# Measurement of magic-wavelength optical dipole trap by using the laser-induced fluorescence spectra of trapped single cesium atoms

**BEI LIU,**[1,2] **GANG JIN,**[1,2] **RUI SUN,**[1,2] **JUN HE,**[1,2,3] **AND JUNMIN WANG** [1,2,3,*]

[1] *State Key Laboratory of Quantum Optics and Quantum Optics Devices, Shanxi University, Tai Yuan 030006, Shan Xi Province, People's Republic of China*

[2] *Institute of Opto-Electronics, Shanxi University, Tai Yuan 030006, Shan Xi Province, People's Republic of China*

[3] *Collaborative Innovation Center of Extreme Optics, Shanxi University, Tai Yuan 030006, Shan Xi Province, People's Republic of China*

\* *wwjjmm@sxu.edu.cn*

**Abstract:** Based on the multi-level model, we have calculated light shifts for Zeeman states of hyperfine levels of cesium (Cs) $6S_{1/2}$ ground state and $6P_{3/2}$ excited state. The magic-wavelength linearly-polarized optical dipole trap (ODT) for Cs $6S_{1/2}|F=4, m_F=+4\rangle$ - $6P_{3/2}|F'=5, m_F=+5\rangle$ transition is experimentally constructed and characterized by using the laser-induced fluorescence spectra of trapped single Cs atoms. The magic wavelength is 937.7 nm which produces almost the same light shift for $6S_{1/2}|F=4, m_F=+4\rangle$ ground state and $6P_{3/2}|F'=5, m_F=+5\rangle$ excited state with linearly-polarized ODT laser beam. Compared to undisturbed Cs $6S_{1/2}|F=4, m_F=+4\rangle$ - $6P_{3/2}|F'=5, m_F=+5\rangle$ transition frequency in free space, the differential light shift is less than 0.7 MHz in a linearly-polarized 937.7 nm ODT, which is less than 1.2% of the trap depth. We also discussed influence of the trap depth and the bias magnetic field on the measurement results.

## 1. Introduction

As a sort of non–classical light source, single-photon sources are important for quantum communication protocols, quantum cryptography protocols as well as linear optics quantum computing [1-3]. Furthermore, photons can be used to code quantum information. Single photon emission has been demonstrated for many different sources. Most of single-photon sources are based on single-emitter quantum systems, such as single atoms [4-7], single ions [8], single quantum-dots [9-10], single molecules [11] and single N/V-centers in diamond [12]. The single atoms, trapped in an optical dipole trap (ODT), is a candidate for single-photon source in principle. It emits a single photon during its spontaneous decay from an excited state to a ground state, and can't emit a second photon until it is re-excited. Some single photon applications require photons to be indistinguishable from one to another. However, in an ODT, the trapping laser induces a light shift on the atom that shifts the transition frequency between the ground and the excited states. This shift is position-dependent and time-dependent due to the thermal motion of the trapped atom, which directly connected to the mean kinetic energy of the atom. The thermal motion of the atom in the trap will explore differential light shift of the transition, which will lead to broadening of the emission spectrum [13] and can thereby reduce the achievable two-photon interference contrast [14].

In order to eliminate the differential light shift, the ODT can be switched off during the excitation/emission processes [5]. However, with this method, the trapping lifetime is reduced, which leads to a lower repetition rate of single photons. The ac Stark shift of the trapping laser can also be eliminated in the blue-detuned dark ODT [15-16]. However, although it is possible to trap atoms, the micrometer-scale blue-detuned dark ODT generally require more complicated experimental setup and not easy to implement. Another alternative method is constructing a magic-wavelength ODT, which can eliminate the differential light shift [17-21]. Using this method, the transition frequency and the emission of the photons are therefore the same as in free space. The specific wavelength of the trapping beam required to achieve zero differential light shift for the concerned transition is called the magic wavelength.

Magic wavelength in neutral atomic systems have been calculated and verified in several experiments [17-21]. In 2003, McKeever et al [18] have calculated the magic wavelength of ODT laser for cesium (Cs) atoms, and constructed a magic-wavelength standing-wave ODT for trapping single atom. In 2010, Phoonthong et al [21] have measured the differential light shift of Cs (F=4) - (F'=3) transition as a function of the trapping wavelength. However, the calculation of the magic wavelengths indicate that the differential light shift is sensitive to the Zeeman state. Thus the trapped atoms can still experience a small light shift. Specifically, considering the different $m_F$ states, we construct a magic-wavelength ODT for Cs $6S_{1/2}$|F=4, $m_F$=+4⟩ - $6P_{3/2}$ |F'=5, $m_F$=+5⟩ transition, which was also measured by using the laser-induced fluorescence (LIF) spectra of trapped single Cs atoms [22]. The fluorescence spectra is obtained by exciting the single atoms with a $\sigma^+$-polarized Cs $6S_{1/2}$ |F=4, $m_F$=+4⟩ - $6P_{3/2}$ |F'=5, $m_F$=+5⟩ probe pulse laser, and collecting the spontaneously emitted photons.

In this paper, we report our experiment to eliminate the differential light shift of Cs $6S_{1/2}$ |F=4, $m_F$=+4⟩ - $6P_{3/2}$ |F'=5, $m_F$=+5⟩ transition by using the magic-wavelength ODT. By using the single atoms, it has eliminate the collision induced spectral broadening and shifting. We measure the differential light shift of Cs $6S_{1/2}$ |F=4, $m_F$=+4⟩ - $6P_{3/2}$ |F'=5, $m_F$=+5⟩ transition as a function of the trapping laser wavelength and analyze the various influence mechanisms of the measurement result.

## 2. Experimental Setup

The schematic of experimental setup is shown in Fig. 1(a). An external-cavity diode laser (ECDL) and tapered amplifier (TA) are used to construct a linearly polarized magic-wavelength ODT. The ECDL can be tuned among 930 ~ 940 nm. The ODT beam is strongly focused by a high numerical aperture (NA = 0.29) objective lens assembly mounted outside the ultra-high vacuum glass cell. The vacuum glass cell has an external dimensions of 30×30×120 mm$^3$. At the focal point, the trapping beam has a $1/e^2$ radius of 1.6 μm. The trap depth is about 3 mK for laser power about 18 mW. The LIF signal of single trapped atom is also collected by the same objective lens and detected by the single-photon-counting module (SPCM, Perkin-Elmer SPCM-AQR15). The P7888 card (two-input multiple-event time digitizers, FAST Com Tech) is used to record the SPCM signal [6]. The quantization axis is defined by a 0.5 Gauss bias magnetic field along the z direction.

The probe beam is derived from the main laser system at 852 nm and locked using polarization spectroscopy. The frequency fluctuation of the probe laser is about ±100 kHz in 200 s [23]. The probe beam is σ$^+$-polarized using a Glan-Taylor prism and a quarter-wave plate, relative to the quantization axis and the $1/e^2$ waist is 12 μm [24]. In order to measure the differential light shift of Cs $6S_{1/2}$ |F=4, $m_F$=+4⟩ - $6P_{3/2}$ |F'=5, $m_F$=+5⟩ cycling transition, the frequency of the probe laser is adjusted by an acousto-optical modulator (AOM).

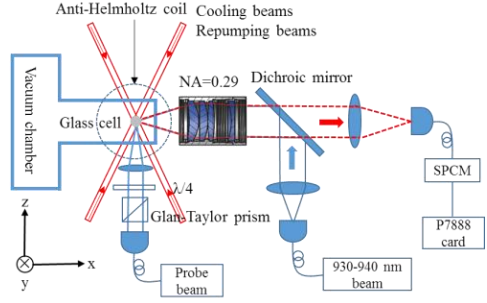

Fig. 1. Experimental setup. The trapped single atoms are excited by a probe beam. The emitted photons are collected by an objective and then coupled into SPCM and P7888 card for measurement.

## 3. Experiment results and discussions

The focused trapping laser beam induces a light shift of the internal states of Cs atoms. Thus the transition frequency between the ground and the excited states are shifted. Taking the multi-level structure of the atom into account, the light shift for the atoms in a linearly polarized ODT is given by [18, 25, 26]:

$$\Delta_{F,m_F} = -3\pi c^2 I(r) \sum_{J'F'm'} \frac{A_{J \to J'}(2J'+1)(2F+1)(2F'+1)}{\omega_{JJ'}^2(\omega_{JJ'}^2 - \omega^2)} \begin{pmatrix} F & 1 & F' \\ m_F & \varepsilon & -m'_F \end{pmatrix}^2 \begin{Bmatrix} J & J' & 1 \\ F' & F & I \end{Bmatrix}^2 \quad (1)$$

where $I(r)$ is the intensity of the trap laser, $\omega_{JJ'}$ is the transition frequency, $\omega$ is the frequency of the trap laser, $A_{J \to J'}$ is the transition rate, also known as the Einstein A coefficient. $\varepsilon$ is the polarization vector, for the linearly polarization ODT $\varepsilon=0$, for the circular polarization ODT $\varepsilon=\pm 1$. The coefficients in round and curly brackets are the 3-j and 6-j symbols. Fig. 2 shows the light shift for each Zeeman levels of $6S_{1/2}$ (F=4) and $6P_{3/2}$ (F'=5) states. In the calculation, the trap depth is about 3 mK. All Zeeman levels of the ground state $6S_{1/2}$ are identical shifted, while the excited states are no longer degenerate. Fig. 2(a) shows that the magic wavelength for Cs $6S_{1/2}$ |F=4, $m_F$=+4⟩ - $6P_{3/2}$ |F'=5, $m_F$=+5⟩ transition of Cs is around 937 nm, and it is different for each Zeeman levels. Fig. 2(b) shows the ratio of the U to Ug, when the ODT laser wavelength is set to 937 nm. U is the light shift for all the Zeeman levels of $6P_{3/2}$ excited state and Ug is the light shift of $6S_{1/2}$ ground state. For $6P_{3/2}$ (F'=5) state, Zeeman levels are split as much as about 15% relative to the ground state shift.

We experimentally construct a magic-wavelength ODT for Cs $6S_{1/2}$ |F=4, $m_F$=+4⟩ - $6P_{3/2}$ |F'=5, $m_F$=+5⟩ transition. In the experiment, the single photon collection efficiency is about 2.1%, the temperature of the trapped atoms is about 60 μk and the lifetime of the atoms in the ODT is about 3.5 s. The magic wavelength is obtained by measuring the differential light shift of Cs $6S_{1/2}$ |F=4, $m_F$=+4⟩ - $6P_{3/2}$ |F'=5, $m_F$=+5⟩ transition. The typical experimental sequence is shown in Fig. 3. The magneto-optical trap (MOT) is loaded from the background vapor for 2 s. After the MOT loading phase, the ODT beam is turned on. After 25 ms MOT and ODT overlap and 10 ms polarization gradient cooling (PGC) phase, the MOT is turned off, and the single atoms are trapped in the ODT [27]. Then a unidirectional σ+-polarized probe laser is used to obtain the LIF spectra of trapped single atoms. During the probing process, the quantization axis magnetic field and repumping laser are always on. In order to detect enough LIF photons and suppress the possibility of losing atoms due to the heating from the probing process. The gated probing/cooling procedure is used and repeated for 500 times [6, 26]. The single atoms is probed for 100 μs and cooled for 900 μs. A low probe saturation parameter s = $I_{probe}/I_{sat}$ ≈ 0.7 is used to avoid remarkable power broadening. A typical duration of 500 ms probing/cooling procedure, the total number of scattered photons $N(t)dt$ can be estimated by [28]:

$$N(t)dt = \frac{\Gamma}{2} s / (1 + 4\left(\frac{\delta - \delta_0}{\Gamma}\right)^2 + s) \times dt \times \eta \qquad (2)$$

where $s$ is the saturation parameter, $\delta$-$\delta_0$ is the detuning of the probe laser which is far from the atomic resonance, $\Gamma/2\pi$ = 5.2 MHz is the natural linewidth, the total effective detection efficiency $\eta$ is about 0.6%. In the detection process, the SPCM is gated to decrease the influence of the background signal. In order to get a large signal to noise ratio, we average over typically 1000 sequences. After averaging over many sequences, the profile of the fluorescence signal measured as a function of the probe detuning $\delta$ should be the Lorentzian profile.

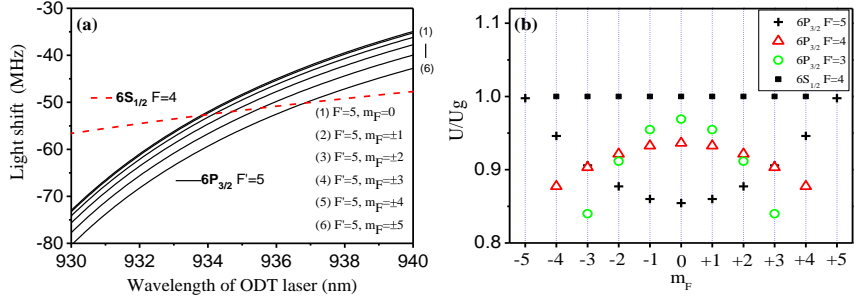

Fig. 2. Light shift for the $6S_{1/2}$ |F=4, $m_F$⟩ and $6P_{3/2}$ |F'=5, $m_F$⟩ state in a linearly polarized ODT. The shifts were calculated by taking the multi-level model into account. (a) The light shift as a function of the wavelength of the ODT laser. The ground states are homogeneous and the excited states shifts are state dependent. (b) In a 937 nm ODT, the ratio of U to Ug as a function of Zeeman states. U is the light shift for all the Zeeman states of $6P_{3/2}$ (F'=3, 4, and 5) excited states and Ug is the light shift of $6S_{1/2}$ (F=4) ground state.

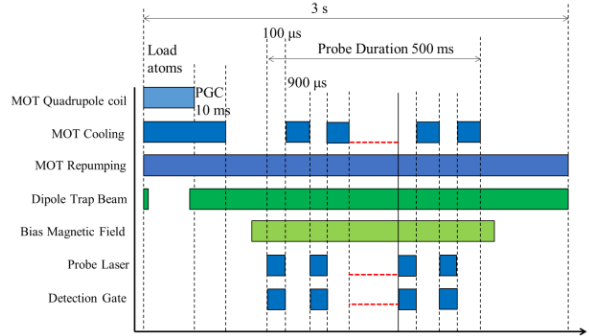

Fig. 3. Experimental sequence. After 10 ms PGC, the single atom is trapped in an ODT. The atom is probed for 100 μs and cooled for 900 μs. In the probing process, the quantum magnetic field is always on. This sequence is typically repeated 1000 times to improve the signal to noise ratio.

Using the technique of LIF spectra of trapped single Cs atoms, we measure the value of the magic wavelength for Cs $6S_{1/2}$ |F=4, $m_F$=+4⟩ - $6P_{3/2}$ |F'=5, $m_F$=+5⟩ transition. As shown in Fig. 4, the LIF spectra are measured for different trap wavelength. In these measurements, the trap beam is linearly polarized and the trap depth is set to ~ 3 mK. The number of detected LIF photons depends on the power and the detuning of the probe laser beam. In Fig. 4(a), the intensity of the probe beam is hold at s ~ 0.7 and the probe frequency is scanned over a range of 80 MHz by tuning the AOM driver frequency. The quantization magnetic field is about 0.5 Gauss. The obtained detected photons for each data is an average over 1000 sequences for the same atom. The data was fitted by Lorentzian profile. The differential light shift is $\delta$ = -16.7 MHz, -0.7 MHz, 5.8 MHz for the trap wavelength of 932.7 nm, 937.7 nm, 940.7 nm, respectively. In Fig. 4(b), we compare the experimental data for the differential light shift as a function of the trap wavelength to the theoretical value as calculated based on multi-level model. The zero differential light shift wavelength is identified as magic wavelength. Our result confirms that the magic wavelength is about 937.7 nm. For this wavelength, we can implement a magic-wavelength ODT with almost zero differential light shift for Cs $6S_{1/2}$ |F=4, $m_F$=+4⟩ - $6P_{3/2}$ |F'=5, $m_F$=+5⟩ transition for the linearly-polarized ODT laser beam. Compared to the undisturbed transition frequency in free space, the frequency shift is less than 0.7 MHz, which is about 1.2% of the ODT depth. We also measure the differential light shift for different trap depths when the trap wavelength is set to 937.7 nm. As shown in Fig. 5, experimental data are obtained by measuring the differential light shift of Cs $6S_{1/2}$ |F=4, $m_F$=+4⟩ - $6P_{3/2}$ |F'=5, $m_F$=+5⟩ transition. For the different trap depth, the probe detuning is ~ 0 MHz. This result shows that the differential light shift does not depend on the trap depth in 937.7 nm magic-wavelength ODT, which further verify our measurement result. One major error of the magic wavelength measurement comes from the power and frequency fluctuation of the probe laser.

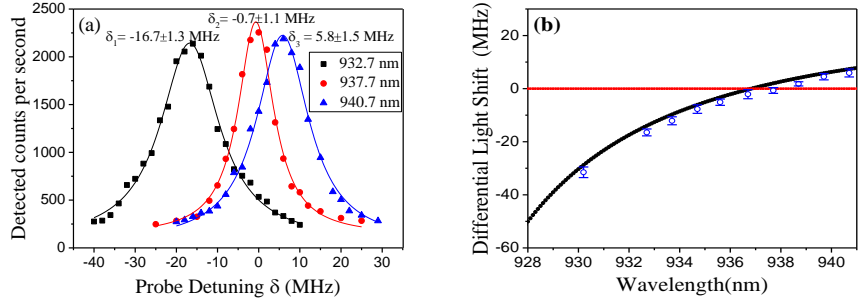

Fig. 4. Measurement of the magic wavelength. (a) LIF spectra of single trapped atoms. Detected photon counts as a function of the detuning of the probe laser for different trap wavelength. The trap depth is ~ 3 mK. Solid lines are Lorentzian profile. (b) The differential light shift versus the trap wavelength. The blue date is the experimental results, the black line is the theoretically expected values. The error bars show the fitting errors of the Lorentzian profile.

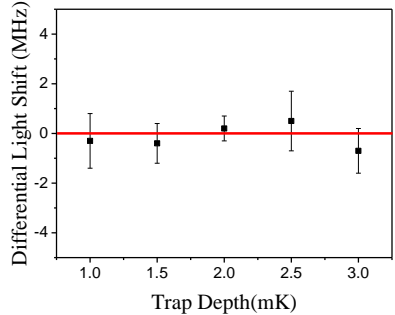

Fig. 5. Differential light shift versus the trap depth. The trap wavelength is 937.7 nm. Each data is obtained by measuring the LIF spectra of single trapped atoms. The differential light shift does not depend on the trap depth in a 937.7 nm ODT. The error bars show the fitting errors of the LIF spectra of trapped single atoms.

The uncertainty of measurement result can be improved by reducing the frequency fluctuation of the probe laser. Otherwise, several processes may limit the uncertainty of measurement result. Firstly, the light shift depends on the polarization of the trap laser beam. Considering the polarization of the ODT laser beam, the polarization-dependent light shift can be derived [28] as follows:

$$\Delta = \frac{\hbar\Gamma^2 I}{8I_{sat}} \left( \frac{1-\varepsilon g_F m_F}{3\Delta_1} + \frac{2+\varepsilon g_F m_F}{3\Delta_2} \right) \quad (3)$$

where $\varepsilon$ is the unit-normalized polarization vector. The ODT laser beam with different polarizations can result in different light shifts, which will influence the value of the magic wavelength. In the experiment, a Glan-Taylor prism which has a typical extinction ratio of ~ $1 \times 10^5 : 1$ is used to create a purely linear polarization of ODT laser beam.

The second important contribution to the shifts of atomic transition frequency is the Zeeman shifted induced by bias magnetic fields. The $6S_{1/2}$ |F=4, $m_F$=+4⟩ ground state experiences a magnetic shift of 0.37 MHz/Gauss and $6P_{3/2}$ |F'=5, $m_F$=+5⟩ excited state experiences a magnetic shift of 0.56 MHz/Gauss. In our experiment, the residual magnetic field is actively stabilized to values of B < 10 mG. So the main influence factors of magnetic fields come from the quantization axis magnetic fields. In Fig. 6, we measure the differential light shift of Cs $6S_{1/2}$ |F=4, $m_F$=+4⟩ - $6P_{3/2}$ |F'=5, $m_F$=+5⟩ transition as a function of the quantization axis magnetic field. Experimental data are obtained by measuring LIF spectra of trapped single atoms. When the magnetic field is about 1.5 Gauss, the differential light shift is ~ 0 MHz.

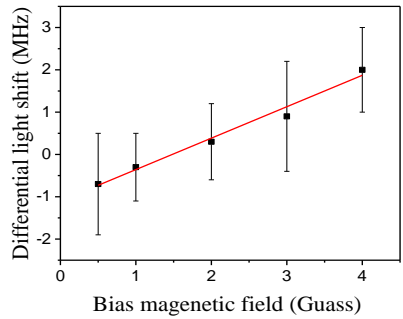

Fig. 6. Differential light shift versus the quantization axis magnetic field. When the magnetic field is about 1.5 Gauss, the differential light shift is ~ 0 MHz. The error bars show the fitting errors of the LIF spectra of trapped single atoms.

## 4. Conclusions

The magic-wavelength ODT is valuable for precision measurement, frequency metrology, and coherent manipulations of quantum systems [29-30]. To conclude, we have constructed a magic wavelength for Cs $6S_{1/2}$ |F=4, $m_F$=+4⟩ - $6P_{3/2}$ |F'=5, $m_F$=+5⟩ cycling transition. In such ODT, the ground state and the excited state have almost the same light shift, which resulting the transition frequency and the emission of the photons are therefore the same as in the undisturbed case in free space. We also have presented a technique to measure the magic wavelength. By using the LIF spectra of single trapped atom, the magic wavelength for Cs $6S_{1/2}$ |F=4, $m_F$=+4⟩ - $6P_{3/2}$ |F'=5, $m_F$=+5⟩ transition has been verified experimentally. The measured value is influenced by the he power and frequency fluctuation of the probe laser, ODT polarization, and Zeeman shift. Our experimental method can also be applied to measure magic wavelength for other different $m_F$ states. In the future, we will perform the two-photon interference experiment to analyze the indistinguishability of the single photons.

## Funding